\documentclass[a4a, 12pt]{article}
\usepackage[margin=1in]{geometry}
\usepackage{graphicx}
\usepackage{placeins}
\usepackage{amssymb}
\usepackage{setspace}
\usepackage{hyperref}
\usepackage[dvipsnames]{xcolor}
\usepackage{color,soul}
\usepackage{float}
\usepackage[justification=centering]{caption}
\usepackage{subcaption}
\usepackage{pdflscape}

\makeatletter
\let\@fnsymbol\@arabic
\makeatother

\begin{document}

\newcommand{\pc}{principal component}
\newcommand{\pcs}{principal components}

\title{The principal components of electoral regimes -- Separating autocracies from pseudo-democracies}
\author{ Karoline Wiesner,$^{1\ast}$ Samuel Bien,$^{1}$ Matthew C. Wilson$^{2}$\\
\\
\normalsize{$^{1}$Institute of Physics and Astronomy, University of Potsdam, Germany}\\
\normalsize{$^{2}$Department of Political Science, University of South Carolina, U.S.}\\
\\
\normalsize{$^\ast$Corresponding author; E-mail:  karoline.wiesner@uni-potsdam.de.}
}

\maketitle\thispagestyle{empty}


\begin{abstract}
A critical issue for society today is the emergence and decline of democracy worldwide. 
It is unclear, however,  how democratic features, such as elections and civil liberties, influence this change. Democracy indices, which are the standard tool to study this question, are based on the a priori assumption that  improvement in any individual feature strengthens democracy overall. 
We show that this assumption does not always hold. 
We use the V-Dem dataset for a quantitative study of  electoral regimes 
worldwide during the 20th century. We find a so-far overlooked trade-off between election quality and civil liberties. In particular, we identify a threshold in the democratisation process at which the correlation between  election quality and civil liberties flips  from negative to positive. 
Below this threshold we can thus clearly separate two kinds of non-democratic regimes: autocracies that govern through tightly controlled elections and regimes in which citizens are free but under less certainty -- a distinction that existing democracy indices cannot make. 
We discuss the stabilising role of election quality uncovered here in the context of the recently observed decline in democracy score of long-standing democracies, so-called `democratic backsliding' or `democratic recession'.
\end{abstract}

\section{Introduction}

By the end of the 20th century, scholars and practitioners alike had come to see democracy not only as universally valuable, but a logical conclusion of political development \cite{Fukuyama1992, sen1999universal}. However, two trends quickly sobered those optimistic views. The first was the observation of autocratic regimes with seemingly democratic institutional features, which some labeled ``competitive authoritarian'' regimes or ``illiberal democracies'' \cite{LevitskyWay:2002, Zakaria:1997}.\footnote{Examples include Azerbaijan, Zimbabwe, and Peru \cite{LevitskyWay:2002}.}  The second was the potential for backsliding among established democracies, for which scholars lack explanations \cite{waldner2018unwelcome, hyde2020democracy}. Countries such as the U.S., India, Nicaragua, Poland, and Hungary are often-cited examples of possible backsliders whose decline became apparent after 2015 \cite{cianetti2020rethinking, markey2022strategic, perello2022changes}. Both trends evidence institutional changes that do not represent wholesale `democratization' or `autocratization' and that may not be clear from continuous measures of democracy that aim to characterize a country's overall democraticness.

Early empirical treatments of democracy tended to represent it in a binary fashion -- e.g., Schumpeter \cite{Schumpeter:1950} and Huntington \cite{Huntington:1991} --, classifying countries as either democratic or not.\footnote{Many scholars still use a binary, categorical approach to representing democracy, such as Cheibub et al. \cite{Cheibubetal:2010} and Boix et al. \cite{Boixetal:2013}, adhering to `minimal' criteria to be considered democratic \cite{Alvarezetal:1996}.}  However, 
the 1990s saw greater efforts to quantify democracy and widespread use of continuous measures. Improvements in data collection and aggregation methods led to indices that combined together scores representing features such as the \emph{extent} of competitiveness and inclusion and additional freedoms (examples include \cite{Arat:1991,CoppedgeReinicke:1990,Hadenius:1992}, for a review, see 
\cite{MunckVerkuilen:2002}.)  

Such indices implicitly make  the assumption that these features correspond together, however, such that an observation with a higher value on the overarching index has subcomponent values that are greater than or equal to those of an observation with a lower score.  The question remains whether ``all things work together for good,'' and whether being stronger on some features over others matters for political development.  This is central to criticisms that composite indices mask important variations in regimes \cite{GleditschWard:1997} and that inconsistencies between different aspects shape countries' prospects for democracy \cite{Dahl:1971,KnutsenNygard2015}.

To help understand how elections and civil liberties in combination  influence democratic development, we examine  the multidimensional dataset provided by the Varieties of Democracy (V-Dem) project, an ambitious coding project of hundreds of variables relating to attributes of democracy. 
 By applying principal component analysis to the subset of the V-Dem data that relates to electoral qualities of democracy, over the period 1900 to 2021, we identify a two-dimensional subspace that contains over 80\% of the variance of the data. The second dimension represents the so-far overlooked trade-off between electoral control and citizen freedom. This allows us to clearly separate electoral autocracies from countries in which citizens are free but which potentially deal with corruption and violence as a result of those freedoms -- a distinction that existing democracy indices cannot make. 
Furthermore, by studying the variables constituting the first and the second component, we discover a threshold in the first component at which the correlation between election quality and civil liberties turns from negative to positive. Only once this threshold is crossed do election quality and civil liberties ''work together for good''.  We discuss the stabilising role of elections uncovered here in the context of the recently observed decline in democracy score of long-standing democracies, so-called ‘democratic backsliding’ or
‘democratic recession’.

\section{Methods}
\label{sec:methods}

\subsection{Data}
\label{subsec:data}

We use the publicly available dataset on electoral democracy, collected and provided by the Varieties of Democracy project (V-Dem).\footnote{We use the latest version (version 12, 2022) of the V-Dem data, \url{https://www.V-Dem.net}.} V-Dem is a large-scale collaborative effort that uses expert coding to generate quantitative data on the quality of various attributes of democracy \cite{vdem-codebook}, for over 190 countries -- dating back decades to centuries, depending on the country -- , with annual resolution. The project involves surveying a large number of country experts and using a Bayesian measurement model to estimate latent values \cite{vdem-methodology}.  The surveys tend to ask respondents to rate the level of openness/strength of an institution, such as election intimidation or media censorship, on an ordinal scale (e.g., ``low," ``intermediate," ``high").  Based on the responses, the measurement model estimates reliability between respondents and generates point estimates for each question.  
The project combines information for different attributes into mid-level indices that represent specific concepts, such as the freedom of expression and election quality, which are also used to represent democracy more generally. 

One of the primary indices that V-Dem publishes is the Electoral Democracy Index (EDI), which aims to represent the combined institutional guarantees suggested by Dahl \cite{Dahl:1971}. The index is based on quantitative information on over forty variables relating to electoral democracy. 
We use the relevant data available for the years 1900 -- 2021, yielding 12,296  data points (i.e. country-year events) in total. In the Supplementary Materials, we give further details on the construction of the EDI and list the questions given to the experts for  each of the 24 variables.

\subsection{Principal Component Analysis}
To evaluate the extent to which there are underlying dimensions in the data, we use principal component analysis (PCA). PCA is one of the most simple and robust techniques for dimensionality reduction. It is one of a family of statistical techniques for representing high-dimensional data on a lower-dimensional linear subspace with as little information loss as possible. The variance retained in the lower-dimensional subspace is a measure of the information that is kept. 
 For easier comparability of the components, we rescaled the values to the interval $[-1, 1]$ while preserving the mean of zero. 
 
\section{Results and Discussion}
\label{sec:results}

\begin{figure}[htbp!]
\centering
\includegraphics[width=\linewidth]
{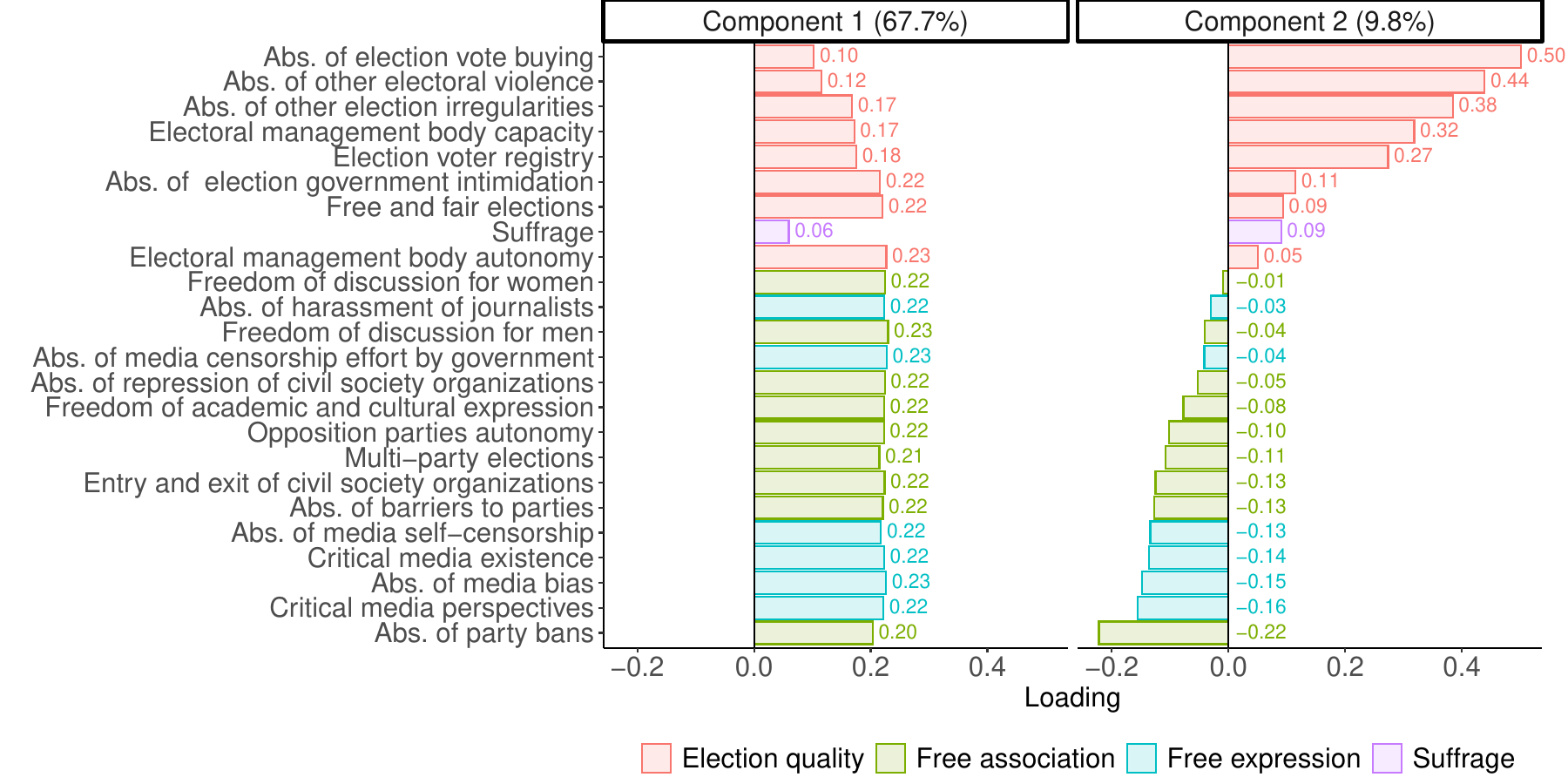}
\caption{Variable loadings on first two principal components. }\label{fig:hwj}
\end{figure}

\begin{figure}[htbp!]
\centering
\begin{tabular}{cc}
\includegraphics[height=.45\textwidth]{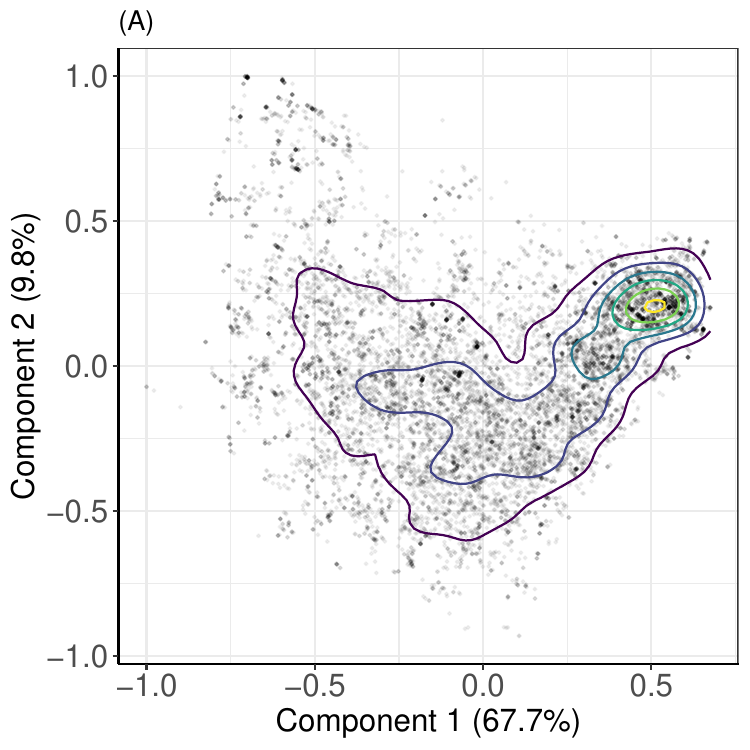}
\includegraphics[height=.45\textwidth]{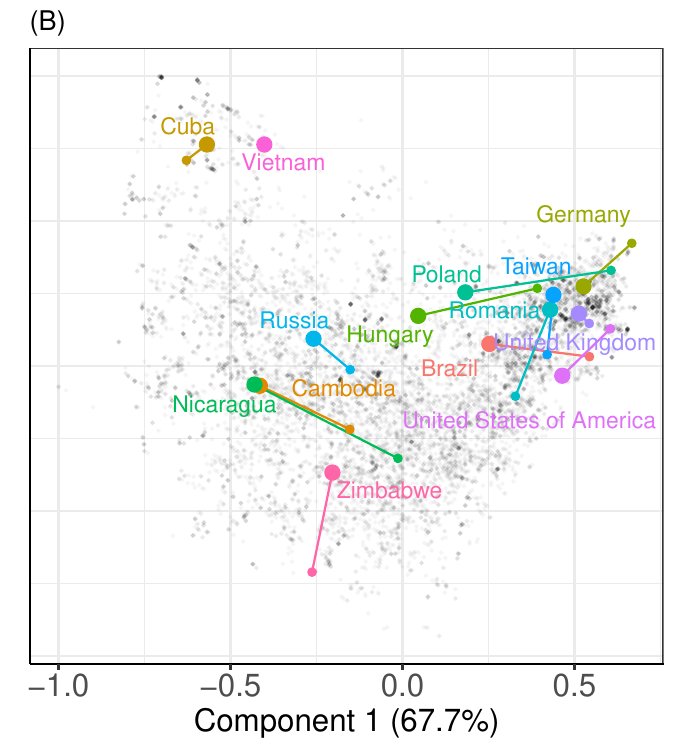}
\end{tabular}
\caption{\textbf{PC2 vs PC1} for all 12,296 data points (correlation $=0$, per definition), (A) with kernel density estimation, (B) with example trajectories for the years 2011 -- 2021 (small dot -- large dot).} \label{fig:plw}
\end{figure}

The main results of the PCA are summarised  in Figs.~\ref{fig:hwj} and \ref{fig:plw}. Fig.~\ref{fig:hwj} depicts how each V-Dem variable loads  onto the first (PC1) and second (PC2) principal component, respectively. Shown in brackets is the variance that is retained by each component: together, the first two {\pcs} account for almost 80 percent of variation in the data. 
Variables are colour-coded according to four classes: `free association', `free expression', `election quality', and `suffrage'. The variable `suffrage' is its own class and  does not contribute significantly to either of the first two principal components. In fact, it predominantly loads onto the third principal component (explained variance: 4.8\%, suffrage loading: 0.85, not shown here). This is possibly due to the fact that it is effectively a discrete variable with dominant values ($0.5$ for male suffrage only and $1.0$ for universal suffrage). In 2021, 98.8\% of countries had universal suffrage. 

By design, a high score in a V-Dem variable indicates `democraticness'  and a low score indicates the lack thereof (see Section~\ref{subsec:data} for details). As seen in Fig.~\ref{fig:hwj}, all V-Dem variables load  positively onto PC1 (i.e. have positive correlations with it), with almost uniformly distributed weight. Hence, ranking countries according to their PC1 score, moving from most negative to most positive values, corresponds to ranking them in increasing democratic quality. 
The fact that the Pearson correlation between PC1 and EDI is very high ($\rho=0.941$) shows that, overall, the EDI is well-designed as a measure of democracy, insofar as the variables tend to move in the same direction. 

PC2, on the other hand, has both positive and negative variable loadings. There is a clear distinction between  variables relating to `free association' and `free expression', which all load negatively onto PC2, and variables relating to election quality and election management capacity -- specifically to the absence of vote buying, election irregularities and violence (i.e. all election variables except \texttt{v2elmulpar}) --, which load positively onto PC2. 
Thus, PC2 potentially represents a trade-off between variables associated with `election quality' (all of which have a positive weight) and variables representing civil liberties (all of which have a negative weight). 
Mathematically, there are (at least) two ways in which a country can score intermediate on PC2 (the original variables take on negative as well as positive values): it either grants its citizens few civil liberties, while maintaining a reasonable level of election quality, or it has a very high level of election quality while granting extensive civil liberties\footnote{In fact, it is a continuous scale, on which the election quality is matched by the lack of civil liberties.}. There is only one way, mathematically speaking, in which a country can score high on PC2: it exhibits a very low level of civil liberties combined with a high level of election quality. We will revisit these cases in the discussion below (see Fig.~\ref{fig:9sj}). 

The association between the first two principal components and the V-Dem variables suggests that PC1 more closely represents `democraticness' associated with electoral competition combined with respect for civil rights and liberties, while PC2 indicates the trade-off between a state's ability to effectively carry out elections and its respect for civil liberties.  This is supported by the observation that some election variables that are particularly important for respecting citizens' preferences, such as the autonomy of election management, election fairness, and absence of intimidation, are more strongly associated with the first component. Notably, the second principal component is very weakly correlated with the EDI ($\rho=0.124$).  

In Fig.~\ref{fig:plw}, all 12,296 country-year data points are plotted according to their PC1-PC2 values. The left-most panel (A) superimposes a standard kernel-density estimate, which shows that the greatest concentration of observations is relatively high on PC1 (0.5) and somewhat lower on PC2 (2.4). Thus, the most populated area is more democratic and with moderately strong electoral control. At the same time, however, there is considerable spread in observations, with some scoring very low on PC1 and very high on PC2.  It is suggestive that some regions of the PC1-PC2 plane are unpopulated. The reason will be become clear in the discussion of Fig.~\ref{fig:9sj} below.  The right-most panel (B) overlays example countries, showing that the upper-left area of the plot is occupied by electoral autocracies such as Cuba and Vietnam, which are characterized by weak civil liberties but that exercise tight control over elections as a tool for legitimation. Based on countries' trajectories between 2011 and 2021 we also see that some, more democratic regimes have moved away from the most populous area, which illustrates backsliding. The relationship between PC1 and PC2 therefore reveals some important information.

\subsection{Election quality vs civil liberties}

\begin{figure}[htbp!]
\centering
\begin{tabular}{cc}
\includegraphics[height=0.45\textwidth]{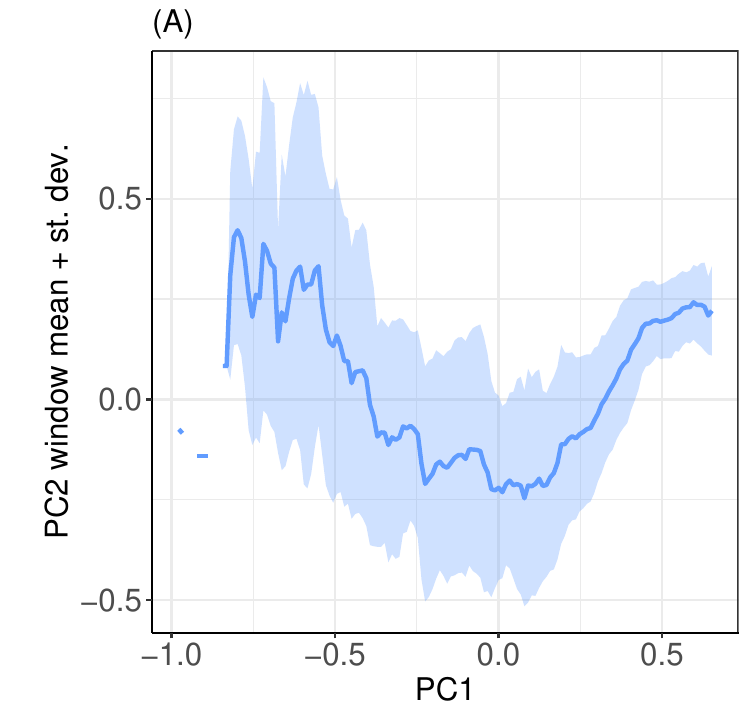} &
\includegraphics[height=.45\textwidth]{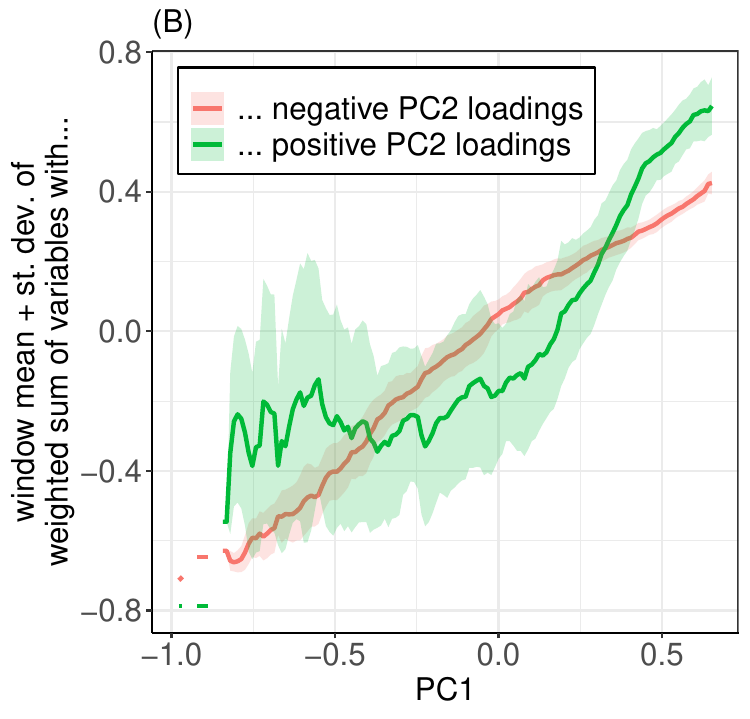}
\end{tabular}
\caption{(A) \textbf{The mean score of PC2  in a sliding window of PC1} with a width of 0.4 and a step size of 0.1. (B)~\textbf{The mean score of a decomposition of PC2} -- 
\\Green line:  (the weighted sum of) variables relating to `election quality', all of which positively correlate with PC2. Red line: (the weighted sum of) variables relating to civil liberties (`free association', `free expression'), all of which negatively correlate with PC2. Thus, the green line minus the red line equals the blue line in (A).}\label{fig:soh}
\end{figure}

To disentangle the two competing contributions to PC2, Figs.~\ref{fig:soh}(A) and (B) graph average values of PC2 against PC1, following a sliding window approach.  In Fig.~\ref{fig:soh}(A), we plot the mean and standard deviation of PC2 against PC1 for all country / year events in a sliding window of width 0.4 and step size 0.1. As PC1 values increase, the average PC2 score first drops and then increases again, with a turning point around PC1 $\approx 0$. The variance in PC2 continuously decreases as one moves toward higher PC1 values, converging to a comparatively narrow distribution for PC1 values above $\approx 0.5$, while the average PC2 score reaches a plateau at around the same point.

To understand this non-linear relation between PC1 and PC2, in Fig.~\ref{fig:soh}(B) we separate the V-Dem variables into those that load negatively onto PC2 (red curve, variables relating to `free association' and `free expression' plus the variable \texttt{v2elmulpar}, which captures the presence of multi-party elections) and those that load positively onto it (green curve, variables relating to `election quality', i.e. all election variables except \texttt{v2elmulpar}).\footnote{A similar technique has been used successfully  to identify scales and thresholds in Holocene social evolution \cite{shin2020scale}.} Here, a clear pattern emerges. For low PC1 values (and therefore also low EDI values), the variables that account for `election quality' (green curve, positive PC2 loadings) greatly fluctuate around an almost constant mean until a turning point around PC1$\approx $0, above which the average of the `clean election' variables continuously increases. The variables that account for `free association' and `free expression'  show a very different behaviour: they continuously increase as PC1 increases, with very narrowly distributed values.  
The turning point in PC2 around PC1 $\approx 0$, which is clearly visible in Fig.~\ref{fig:soh}(A), is revealed as the point at which the quality of elections begins to rise above an apparently minimal value  required for (semi-)democratic regimes. Thus, countries that score low on PC1 include non-democracies that exhibit well-coordinated elections but that do not allow civil liberties, while countries that score higher on PC1 tend to do both. The dip in PC2 that occurs as one moves from low to high values of PC1 indicates the electoral quality attributes being overshadowed by civil liberties, which illustrates a potential trade-off between election control and freedom that liberalizing regimes face. We interpret this dip as the point in democratic development at which feedback between election quality and civil liberties turns from negative to positive. In other words, in the region PC1 $> 0$, election quality and civil liberties are mutually enhancing, while in the region PC1 $< 0$, election quality and civil liberties can be mutually suppressing.\\

\begin{figure}[htbp!]
    \centering
    \begin{tabular}{cc}
        \includegraphics[height=.45\linewidth]{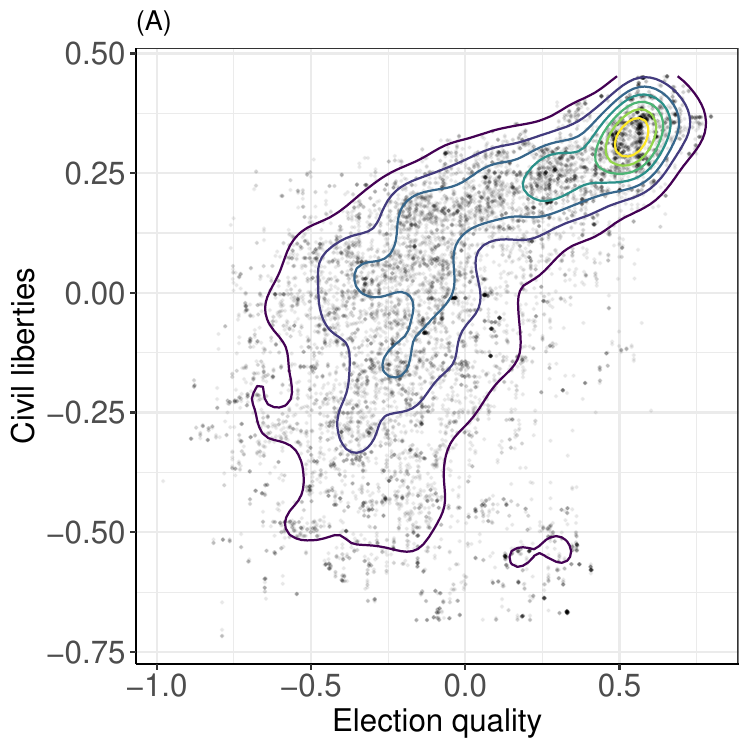} &
        \includegraphics[height=.45\linewidth]{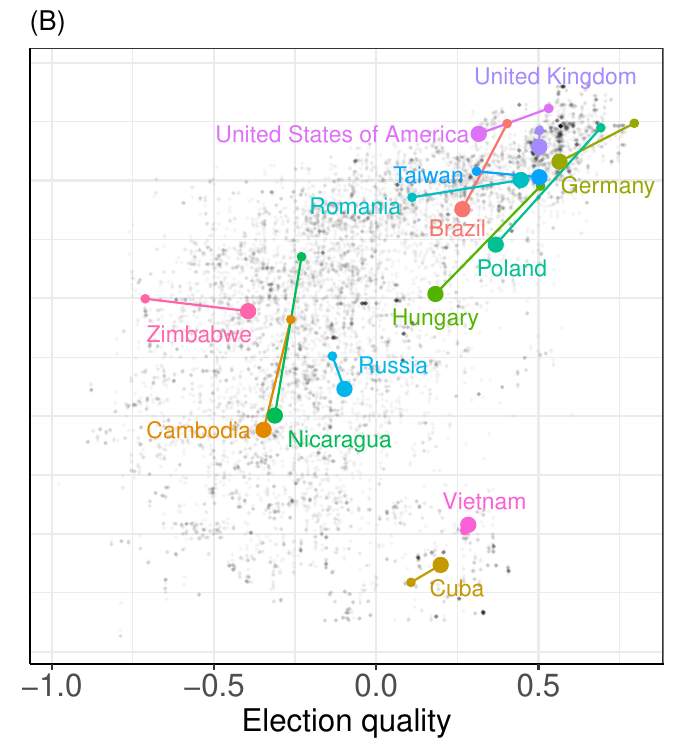}
        \end{tabular}
    \caption{\textbf{Aggregate positive vs. negative PC2 loadings} (i.e. `election quality' vs. `civil liberties') for all 12,296 data points, (A) with kernel density estimation, (B) with example trajectories for the years 2011 -- 2021 (small dot -- large dot).}
    \label{fig:9sj}
\end{figure}

In Fig.~\ref{fig:9sj} we compare the two variable groups, with negative and positive loadings onto PC2, respectively, against each other. Labels of selected country positions in the years 2011 and  2021 are shown. 
The highest density of points -- in the upper right -- is that of countries with high levels of civil liberties and of election quality. These are generally considered to be democracies. In general, those countries with the greatest respect for civil liberties also have quality elections. As the level of control over elections decreases,  civil liberties also decrease in general. At the lowest level of civil liberties (civil-liberties values between $-0.5$ and $-1.0$), a broad range of values for `election quality' are observed. The labels of selected country examples strongly suggest that countries with positive `election quality' values and negative civil-liberties values are electoral autocracies, which are known for holding elections but closely limiting opposition. 

Overall, the data in Fig.~\ref{fig:9sj} show that more democratic regimes are more likely to be able to effectively carry out elections, but that some very non-democratic countries -- which are quite low on the first component -- are also associated with capable elections. This observation is well-known in research on autocracies but has never before been demonstrated using continuous measures of democracy.

\begin{figure}[htbp!]
\centering
\begin{tabular}{cc}
    \includegraphics[height=0.45\textwidth]{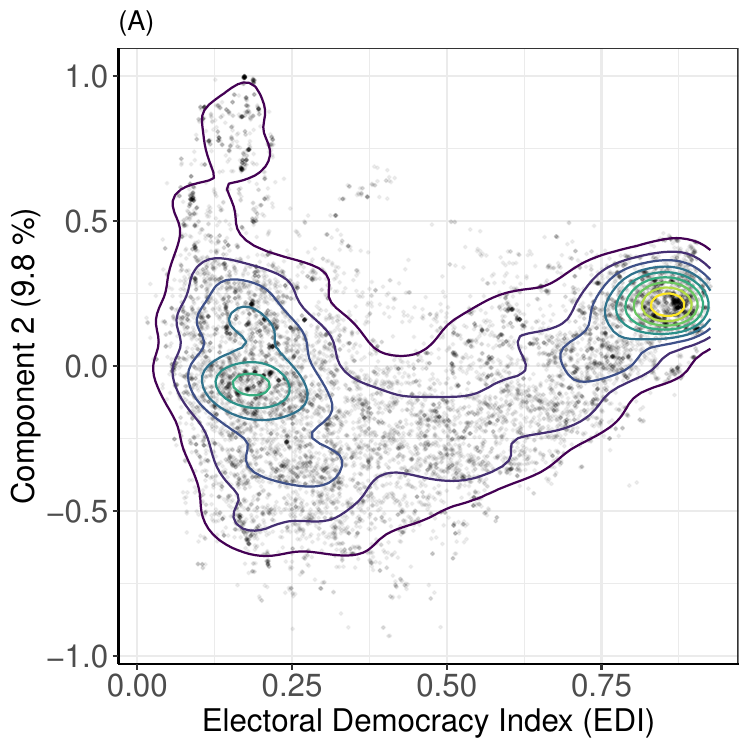} &
    \includegraphics[height=0.45\textwidth]{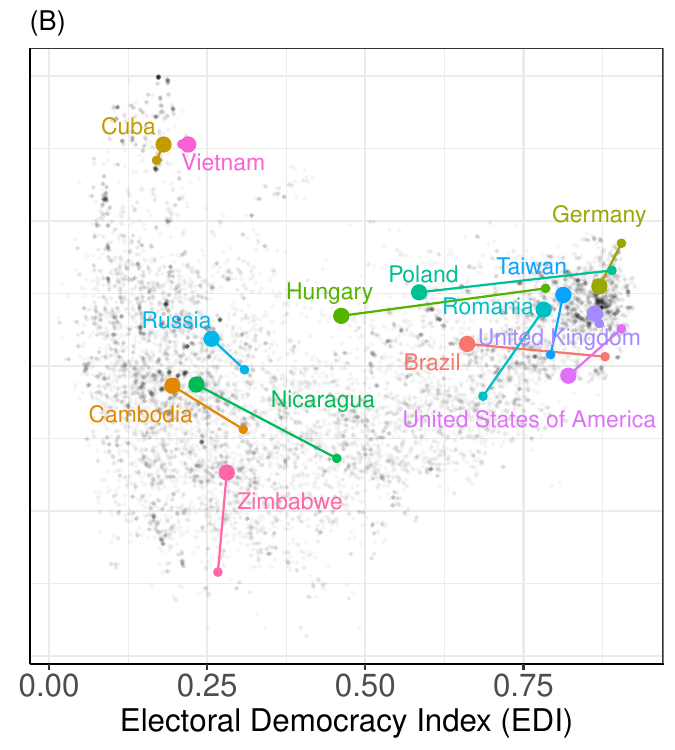} 
\end{tabular}
\caption{\textbf{PC2 vs EDI} for all 12,296 data points (Correlation $=~0.124$), (A) with kernel-density estimate superimposed, (A) with examples for 2011 -- 2021 (small dot  -- large dot).}\label{fig:gdw}
\end{figure}

Figure~\ref{fig:gdw} plots countries' PC2 vs  EDI values. The resulting distribution is very similar to that in Fig~\ref{fig:plw} (showing PC2 vs PC1), since EDI and PC1 are highly correlated ($\rho = 0.941$). Clearly visible is a high variance in PC2 for EDI values $< 0.5$. This indicates that differences between regime type among the non-democratic countries (such as electoral authoritarian and liberalizing regimes) are not resolved by the EDI but become visible in the second principal component. Importantly, countries that score high on the EDI tend to coalesce around similar values of PC2, but there is considerable variation at the lower end of the index that is associated with election quality. 
This is corroborated by specific examples of countries with EDI values around $0.25$ -- a group that includes electoral autocracies (Cuba, Vietnam), failed states (Cambodia, Zimbabwe), and countries like Russia, which are considered to be in the process of solidifying an electoral authoritarian regime. 

Comparing the change in EDI with the change in the positive and negative contributions to PC2 further reveals some interesting difference.\footnote{It is worth keeping in mind, that the relative contribution of PC2 to the variance of the data is 10\%.} 
One group of countries shows almost no change in their EDI value between 2011 and 2021, namely the U.K., Taiwan, and Zimbabwe (Fig.~\ref{fig:gdw}). However, they show significant change in their positive and negative contribution to PC2: The U.K. has declined in its level of civil liberties, with the election quality remaining unchanged, while Taiwan and Zimbabwe have seen an increase in  election quality while civil liberties have remained almost unchanged (on a  high and low level, respectively, see Fig.~\ref{fig:9sj}). Other countries such as the United States have decreased significantly in all three values, as have Hungary and Poland.  

This finding may help shed light on the quality of change captured in the term `democratic backsliding', which has been applied to countries beset by greater political polarization and efforts by dominant parties to undermine the quality of elections under relatively free and fair regimes. 
To summarise, the non-linear relationship between the second principal component and democracy (EDI) 
suggests that -- by exploring variation in the attributes thought to represent democracy -- it picks up a potentially meaningful quality of political regimes and regime change that is obscured by the single dimension represented by the EDI. 

\section{Conclusions}

Our results support three important conclusions. 
First, \emph{the hidden dimension of democracy (PC2) uncovers an important distinction between countries with a similar democracy score (EDI and PC1) but differing levels of election quality and civil liberties.} Evaluating democratic improvements based solely on the EDI may therefore disguise changes in countries that are adopting a more stable form of autocracy by seeking to carry out controlled elections while also not respecting civil liberties. A good example of this is Zimbabwe, which saw a considerable increase on election quality and a decrease on civil liberties between 2011 and 2021 but did not show much change in its EDI score. 

Second, \emph{contrary to assumptions that election quality and civil liberties generally improve together, we identified a threshold in the first dimension at which the correlation between election quality and civil liberties flips from negative to positive.} Here, the interaction between these features of democracy appears to change from mutually enhancing to mutually suppressing.   This finding suggests that there may be a trade-off between the order provided by \textit{controlled} elections and the liberty that is associated with democracy.  This also  helps to explain why some areas of the two-dimensional space tend not to be occupied. 
 
Third, \emph{the quality of elections is a stabilising component of both democracies and (electoral) autocracies.}  The many instances of countries with a high PC2 value and a low EDI (or PC1) value, which we identified as electoral autocracies, suggests that election quality might play a stabilising role in electoral autocracies, despite the lack of civil liberties. However, it begs the question to what extent a decline in election quality has a destabilising effect. Such an effect might have importance for describing `backsliding' among democracies which we lack theories to explain \cite{waldner2018unwelcome}. 
The examples of Hungary, Poland, and the United States of America (Figs.~\ref{fig:9sj} and \ref{fig:gdw}) suggest that the process of `democratic backsliding' is accompanied (or even driven) by a greater decrease in election quality than in the civil liberties. Notably, this differs from backsliding in other countries, such as Brazil and Nicaragua, which may represent different processes.

In view of these results, the question ``which features strengthen or weaken democratic development" seems ill posed. The appropriate question is perhaps ``how do  features interact and, as a result, influence prospects of democratic development". The latter question captures more of the complexity of the concept of democracy \cite{eliassi2020science, wiesner2018stability}.

\section*{Acknowledgments}
K.W. and M.C.W. would like to thank Staffan Lindberg from the V-Dem Institute for making the authors aware of each other's existence. M.C.W. would also like to acknowledge past collaborative efforts with Vanessa Boese-Schlosser that stimulated the present paper.

\textbf{Author contributions:} K.W. and M.C.W. conceptualized the study and conducted the foundational analysis. All authors contributed to the data analysis and to the discussion of the results. 
M.C.W. and S.B. produced the figures. K.W. and M.C.W. wrote the manuscript. 

\textbf{Funding:} S.B. acknowledges the Research Focus Data-Centric Sciences of the University of
Potsdam for support. 

\textbf{Competing interests:} The authors declare that they have no competing interests. 

\textbf{Data and materials availability:} All data needed to evaluate the conclusions in the paper are present in the paper and/or publicly available online.

\newpage\clearpage
\singlespace\raggedright
\bibliographystyle{plain}
\bibliography{Bib}

\begin{thebibliography}{10}

\bibitem{Alvarezetal:1996}
Michael Alvarez, Jose Cheibub, Fernando Limongi, and Adam Przeworski.
\newblock Classifying political regimes.
\newblock {\em Studies in Comparative International Development}, 31(2):1--37,
  1996.

\bibitem{Arat:1991}
Zehra~F. Arat.
\newblock {\em Democracy and Human Rights in Developing Countries}.
\newblock Boulder, CO: Lynne Rienner, 1991.

\bibitem{Boixetal:2013}
Carles Boix, Michael~K. Miller, and Sebastian Rosato.
\newblock A complete data set of political regimes, 1800-2007.
\newblock {\em Comparative Political Studies}, 46(12):1523--1554, 2013.

\bibitem{Cheibubetal:2010}
Jose~Antonio Cheibub, Jennifer Gandhi, and James~Raymond Vreeland.
\newblock Democracy and dictatorship revisited.
\newblock {\em Public Choice}, 143:67--101, 2010.

\bibitem{cianetti2020rethinking}
Licia Cianetti, James Dawson, and Se{\'a}n Hanley.
\newblock {\em Rethinking'democratic Backsliding'in Central and Eastern
  Europe}.
\newblock Routledge, 2020.

\bibitem{vdem-codebook}
Michael Coppedge, John Gerring, Carl~Henrik Knutsen, Staffan~I. Lindberg, Jan
  Teorell, David Altman, Michael Bernhard, Agnes Cornell, M.~Steven Fish, Lisa
  Gastaldi, Haakon Gjerl{\o}w, Adam Glynn, Sandra Grahn, Allen Hicken, Katrin
  Kinzelbach, Kyle~L. Marquardt, Kelly McMann, Valeriya Mechkova, Pamela
  Paxton, Daniel Pemstein, Johannes~von R\"{o}mer, Brigitte Seim, Rachel
  Sigman, Svend-Erik Skaaning, Jeffrey Staton, Eitan Tzelgov, Luca Uberti,
  Yi-ting Wang, Tore Wig, and Daniel. Ziblatt.
\newblock V-dem codebook v12.
\newblock Technical report, Varieties of Democracy (V-Dem) Project, 2022.

\bibitem{vdem-methodology}
Michael Coppedge, John Gerring, Carl~Henrik Knutsen, Staffan~I Lindberg, Jan
  Teorell, Kyle~L Marquardt, Juraj Medzihorsky, Daniel Pemstein, Nazifa
  Alizada, Lisa Gastaldi, et~al.
\newblock V-dem methodology v12.
\newblock {\em V-Dem Working Paper}, 2022.
\newblock \url{https://www.v-dem.net/documents/2/methodologyv12.pdf}.

\bibitem{CoppedgeReinicke:1990}
Michael Coppedge and Wolfgang~H. Reinicke.
\newblock Measuring polyarchy.
\newblock {\em Studies in Comparative International Development}, 25(1):51--72,
  1990.

\bibitem{Dahl:1971}
Robert~A. Dahl.
\newblock {\em Polyarchy: Participation and Opposition}.
\newblock New Haven, CT: Yale University Press, 1971.

\bibitem{eliassi2020science}
Tina Eliassi-Rad, Henry Farrell, David Garcia, Stephan Lewandowsky, Patricia
  Palacios, Don Ross, Didier Sornette, Karim Th{\'e}bault, and Karoline
  Wiesner.
\newblock What science can do for democracy: a complexity science approach.
\newblock {\em Humanities and Social Sciences Communications}, 7(1):1--4, 2020.

\bibitem{Fukuyama1992}
Francis Fukuyama.
\newblock {\em The End of History and the Last Man}.
\newblock New York: Free Press, 1992.

\bibitem{GleditschWard:1997}
Kristian~S. Gleditsch and Michael~D. Ward.
\newblock Double take: A reexamination of democracy and autocracy in modern
  polities.
\newblock {\em Journal of Conflict Resolution}, 41(3):361--383, 1997.

\bibitem{Hadenius:1992}
Axel Hadenius.
\newblock {\em Democracy and Development}.
\newblock Cambridge University Press, 1992.

\bibitem{Huntington:1991}
Samuel~P. Huntington.
\newblock {\em The Third Wave: Democratization in the Late Twentieth Century}.
\newblock University of Oklahoma Press, Norman, OK and London, 1991.

\bibitem{hyde2020democracy}
Susan~D Hyde.
\newblock Democracy’s backsliding in the international environment.
\newblock {\em Science}, 369(6508):1192--1196, 2020.

\bibitem{KnutsenNygard2015}
Carl~Henrik Knutsen and H\aa vard Nyg\aa~rd.
\newblock Institutional characteristics and regime survival: Why are
  semi--democracies less durable than autocracies and democracies?
\newblock {\em American Journal of Political Science}, 59(3):656--670, 2015.

\bibitem{LevitskyWay:2002}
Steven Levitsky and Lucan~A. Way.
\newblock The rise of competitive authoritarianism.
\newblock {\em Journal of Democracy}, 13(2):51--65, 2002.

\bibitem{markey2022strategic}
Daniel Markey.
\newblock The strategic implications of india's illiberalism and democratic
  erosion.
\newblock {\em Asia Policy}, 29(1):77--105, 2022.

\bibitem{MunckVerkuilen:2002}
Gerardo~L. Munck and Jay Verkuilen.
\newblock Conceptualizing and measuring democracy: Evaluating alternative
  indices.
\newblock {\em Comparative Political Studies}, 35(1):5--34, 2002.

\bibitem{perello2022changes}
Lucas Perell{\'o} and Patricio Navia.
\newblock Changes in support for nicaragua’s sandinista national liberation
  front during democratic backsliding.
\newblock {\em Politics}, 42(3):426--442, 2022.

\bibitem{Schumpeter:1950}
Joseph Schumpeter.
\newblock {\em Capitalism, Socialism, and Democracy}.
\newblock New York: Harper, 1950.

\bibitem{sen1999universal}
AK~Sen.
\newblock The universal value of democracy.
\newblock {\em Journal of Democracy}, 10:3--17, 1999.

\bibitem{shin2020scale}
Jaeweon Shin, Michael~Holton Price, David~H Wolpert, Hajime Shimao, Brendan
  Tracey, and Timothy~A Kohler.
\newblock Scale and information-processing thresholds in holocene social
  evolution.
\newblock {\em Nature communications}, 11(1):2394, 2020.

\bibitem{waldner2018unwelcome}
David Waldner and Ellen Lust.
\newblock Unwelcome change: Coming to terms with democratic backsliding.
\newblock {\em Annual Review of Political Science}, 21(1):93--113, 2018.

\bibitem{wiesner2018stability}
Karoline Wiesner, Alvin Birdi, Tina Eliassi-Rad, Henry Farrell, David
  Garc{\'\i}a, Steve Lewandowsky, Patricia Palacios, Don Ross, Didier Sornette,
  and Karim Th{\'e}bault.
\newblock Stability of democracies: a complex systems perspective.
\newblock {\em European Journal of Physics}, 40(1):014002, 2018.

\bibitem{Zakaria:1997}
Fareed Zakaria.
\newblock The rise of illiberal democracy.
\newblock {\em Foreign Affairs}, 76(6):22--43, 1997.

\end{thebibliography}

\newpage

\section*{Supplementary Material for `The hidden dimension in democracy' by 
Wiesner, Bien, Wilson}

The EDI is created by aggregating together five items: an index of the extent to which officials are elected (\textit{v2x\_elecoff}), indices for the freedoms of expression and association  (\textit{v2x\_freexp\_altinf} and \textit{v2x\_frassoc\_thick}), an index representing the quality of elections (\textit{v2xel\_frefair}) and an estimate of the share of the population with suffrage (\textit{v2x\_suffr}).  The aggregation method used by V-Dem is a compromise between a multiplicative and an additive approach.\footnote{The compromise involves multiplying the mid-level indices together and adding weighted values of them together, both of which receive a weight of 0.5.  For more information, refer to \cite{vdem-codebook}.}

In all, the EDI comprises information on approximately 45 different measures: 20 related to elected officials; 9 pertaining to free expression; 6 components of the freedom of association; 8 attributes for election quality, and suffrage.\footnote{Variables for the freedom of association index and election quality index are measured based on whether the regime is coded as an electoral regime according to \textit{v2x\_elecreg}.}  Because  all of the measures that make up the elected officials index are binary variables (e.g., whether the head of state is directly elected or not), we focus our analysis on the 24 remaining indicators (omitting whether the regime is an electoral regime, which is also binary). Details on these 24 variables are provided in the Supplementary Materials. 
Our analysis involves examining variation within the remaining components that compose the indices for free association and expression, election quality, and suffrage. Because election variables are only coded for the years in which there were elections,\footnote{These are \textit{v2elrgstry}, \textit{v2elvotbuy}, \textit{v2elirreg}, \textit{v2elintim}, \textit{v2elpeace}, \textit{v2elfrfair}, and \textit{v2elmulpar}.} we filled in missing values for up to five years between elections.  We also omitted missing values through listwise deletion, ensuring that there are values for each variable for all observations in the sample.

Listed below are the questions given to the experts for  each of the variables. For further details, see \cite{vdem-codebook}. 
Note, that the answer scores are formulated such that a high score reflects a higher quality of `democraticness´. 
In order to clarify the meaning of the second \pc, we list first the definitions of those variables that correlate positively with the second \pc and then those that correlate negatively. All variables correlate positively with the first \pc.

\begin{itemize}     \setlength{\itemindent}{3em}
\item[\textbf{v2elembaut}] 
Does the Election Management Body (EMB) have autonomy from government to apply election laws and administrative rules impartially in national elections?
    \item[\textbf{v2elembcap}] 
    Does the Election Management Body (EMB) have sufficient staff and resources to ad- minister a well-run national election?
    \item[\textbf{v2elrgstry}] In this national election, was there a reasonably accurate voter registry in place and was it used?
    \item[\textbf{v2elvotbuy}] In this national election, was there evidence of vote and/or turnout buying?
    \item[\textbf{v2elirreg}] In this national election, was there evidence of other intentional irregularities by incumbent and/or opposition parties, and/or vote fraud?
    \item[\textbf{v2elintim}] In this national election, were opposition candidates/parties/campaign workers subjected to repression, intimidation, violence, or harassment by the government, the ruling party, or their agents?
    \item[\textbf{v2elpeace}] In this national election, was the campaign period, election day, and post-election process free from other types (not by the government, the ruling party, or their agents) of violence related to the conduct of the election and the campaigns (but not conducted by the government and its agents)? 
    \item[\textbf{v2elfrfair}] Taking all aspects of the pre-election period, election day, and the post-election process into account, would you consider this national election to be free and fair?
\end{itemize}

\noindent We list here the definitions of those variables that correlate negatively with the second \pc. For further details, see \cite{vdem-codebook}.
\begin{itemize}
   \setlength{\itemindent}{3em}
  \item[\textbf{v2meslfcen}] 
  Is there self-censorship among journalists when reporting on issues that the government considers politically sensitive?
   \item[\textbf{v2mebias}] 
   Is there media bias against opposition parties or candidates?
   \item[\textbf{v2mecrit}] 
   Of the major print and broadcast outlets, how many routinely criticize the government?
   \item[\textbf{v2merange}] 
   Do the major print and broadcast media represent a wide range of political perspectives?
   \item[\textbf{v2psparban}] 
   Are any parties banned?
   \item[\textbf{v2psbars}] 
   How restrictive are the barriers to forming a party?
   \item[\textbf{v2psoppaut}] 
   Are opposition parties independent and autonomous of the ruling regime?
   \item[\textbf{v2elmulpar}] 
   Was this national election multiparty?
   \item[\textbf{v2cseeorg}] 
   To what extent does the government achieve control over entry and exit by civil society organizations (CSOs) into public life?

\end{itemize}

\end{document}